\documentclass{article}
\usepackage{graphicx} % Required for inserting images
\usepackage{authblk}
\usepackage{nameref}
\usepackage{setspace}
\usepackage[letterpaper,top=2cm,bottom=2cm,left=3cm,right=3cm,marginparwidth=1.75cm]{geometry}
\title{Mixtures and grain boundaries of P, D, and G triply periodic minimal surfaces}
\author[1]{Chern Chuang}
\author[2]{Bih-Yaw Jin}
\affil[1]{Department of Chemistry and Biochemistry, University of Nevada, Las Vegas, Las Vegas, NV 89154, USA
}
\affil[2]{Department of Chemistry,
National Taiwan University, Taipei, Taiwan}
\date{September 2024}

\begin{document}

\maketitle
\section*{Abstract}
We introduce a square tiling/tetragonal strip representation to the P, D, and G triply periodic minimal surfaces. This approach is useful in identifying mixtures and grain boundaries of these surfaces that might be useful for material sciences or advanced manufacturing purposes. Generalizations to more complicated strip topology (multi-strand) as well as other regular and semi-regular tilings are discussed. Examples of these include double diamond, double gyroid, and triangular P/D/G surfaces. 

\section*{Introduction}
Minimal surfaces are those with surface area minimized locally and consequently are of interest in material science, particularly when interface is concerned, e.g. soap films. Among them, triply periodic minimal surfaces (TPMS) consistent with three-dimensional lattices and crystallographic symmetries are distinctly interesting and find many examples in both synthetic materials as well as biological ones in nature.\cite{LanguageofShape,ChemSocRev2017Amabilino,NaturePhotonics2013Lu,MACKAY1991,GyroidPhotonics,Thomas1988} These TPMS divide space into two interpenetrating networks, which could manifest as amphiphilic surfactant molecules dividing a water phase and an oil phase, photonic crystals with nontrivial band structures, or even finding applications in additive manufacturing.\cite{MASKERY2018220}

The P (for primitive) and D (for diamond) TPMS first discovered in the nineteenth century by Hermann Schwarz and followed by Edvard Neovius. In 1970 Alan Schoen published his discovery of gyroid (G) surface among other TPMS.\cite{Schoen1970} Their occurrences in biology and materials have stimulated much interest in physical science. These three surfaces are all of genus three and of cubic lattice symmetry, and that they can be related by Weierstrass–Enneper parametrization using the same Weierstrass function but with different Bonnet angles, \cite{Schoen1970,Lidin1990} therefore they are often brought up together as popular examples of TPMS.

Due to their three-dimensional nature, it is challenging to visualize and compare these TPMS with different topologies and geometries. This becomes particularly challenging when considering mixtures of or grain boundaries between different TPMS, which are of interest in materials science and macroscopic manufacturing. This work aims to bridge this gap by providing an intuitive construction scheme to analyze the three canonical cubic P, D, and G TPMS. It is shown that by dividing these cubic TPMS into tetragonal cells in a planar square tiling, their geometric similarity to catenoid, helicoid, and certain catenoid/helicoid intermediate, respectively, becomes clear, in relation to the Bonnet transformation that modifies the phase term in their Weierstrass construction.\cite{Schoen1970,Lidin1990} The open boundaries of these tetragonal strips form parallel spirals, making it intuitive to construct mixtures of them or grain boundary between, for example, a D phase and a G phase. The usefulness of this method is further demonstrated with two additional examples: a hexagonal version of gyroid and the double gyroid surfaces.

\section*{Tetragonal unit cells of P, D, and G TPMS}
\begin{figure}
    \centering
    \includegraphics[width=0.7\linewidth]{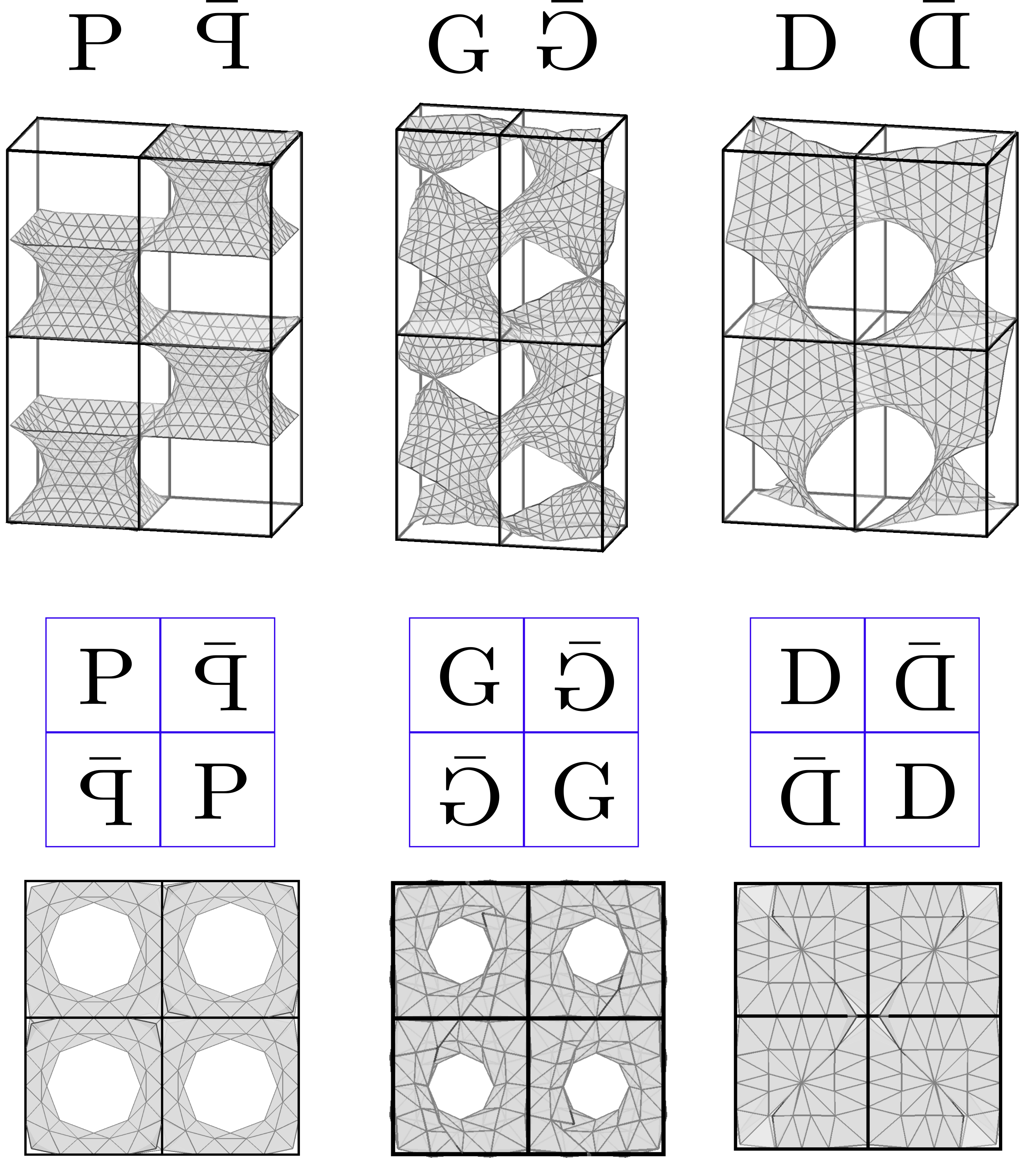}
    \caption{The tetragonal strip/square tiling representations of P/D/G surfaces, side views (upper row) and top views (lower row).}
    \label{fig:fig01}
\end{figure}

This study is based on the observation that all three of the canonical TPMS: the P, D, and G surfaces can be dissected into catenoidal/helicoidal strips that are the unit cells of the underlying tetragonal lattice, a symmetry subgroup of their full cubic symmetry.\cite{BridgesPDG} These strips are related to their nearest neighbors on the lattice by glide reflection. See Fig.~\ref{fig:fig01} for illustration. In (a) we show four neighboring tetragonal unit cells (2-by-1-by-2) in each cases to illustrate their catenoidal/helicoidal nature. Each unit cell is translationally invariant in the $z$-direction, and is related to its $x$- and $y$-neighbors by a glide plane symmetry. To better see the glide symmetry, in (b) we show the top view of four tetragonal unit cells highlighting the glide symmetry (bottom row), symbolically represented as their Roman characters flipped (mirror symmetry) and barred (translation in $z$). This representation will be very useful for the forthcoming sections. 

\section*{Mixed surfaces}
With the tetragonal representations of P, D, and G surfaces defined. We are in a position to discuss hybrid and other generalizations of these surfaces.

\subsection*{Mixed P/D/G}
\begin{figure}
    \centering
    \includegraphics[width=0.9\linewidth]{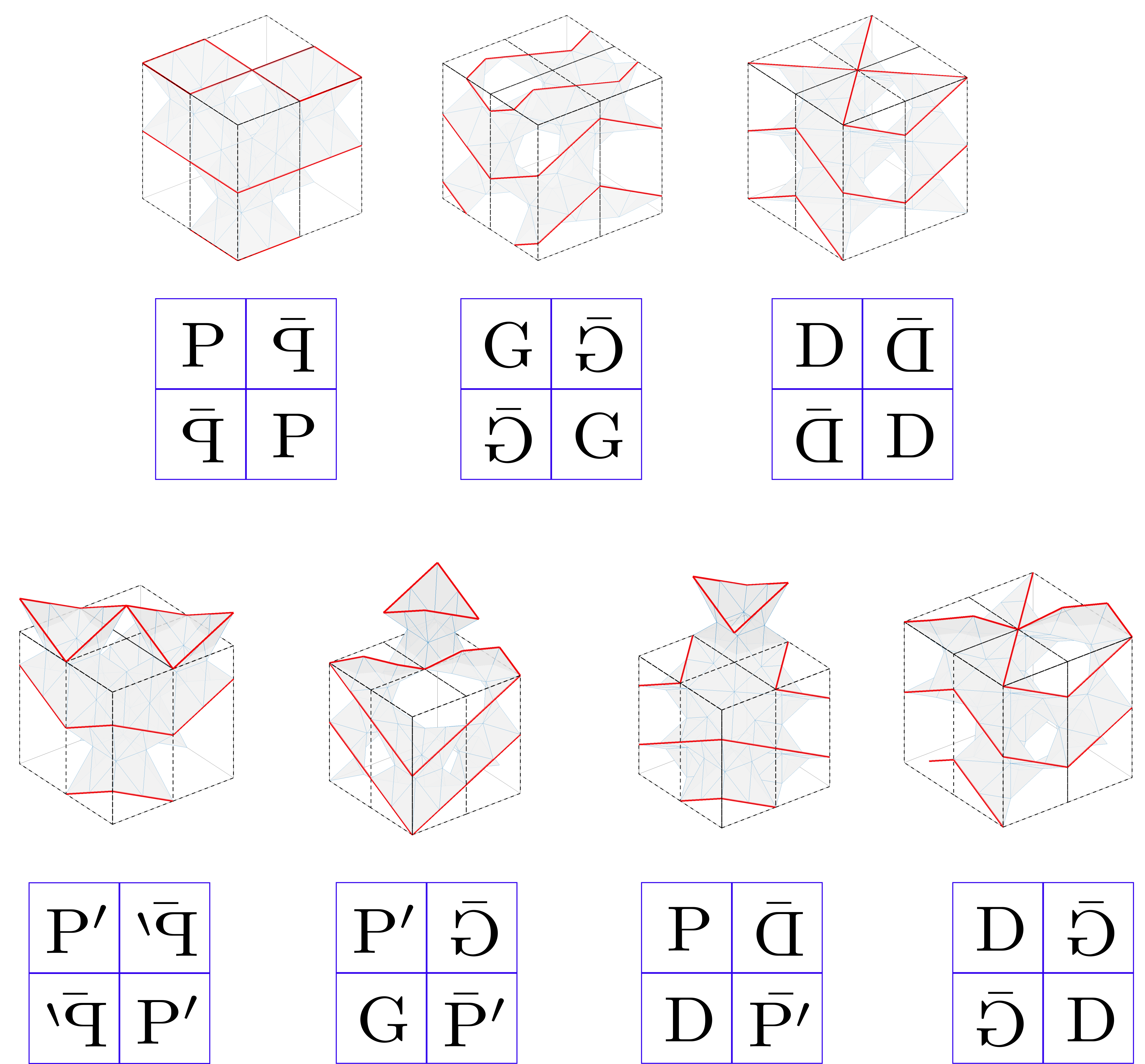}
    \caption{The boundary conditions of the tetragonal strips (upper row) and mixtures of P/D/G surfaces based on these strips (lower row).}
    \label{fig:fig02}
\end{figure}

From Fig.~\ref{fig:fig01}(a) it is readily seen that the lateral boundaries (in the $x$- and $y$-directions) of the tetragonal unit cells of D and G TPMS are two helical broken curves that spiral up the cells and separate equidistantly in the $z$ direction. These are highlighted in Fig.~\ref{fig:fig02}(a). The nearest neighboring cells have opposite handiness to match at the boundaries. Consequently, These two surfaces can be broken down and rejoined (up to a scaling factor of the tetragonal unit cell aspect ratio) seamlessly and form a hybrid D/G surface, shown in the far right of Fig.~\ref{fig:fig02}(b). It is worth mentioning that for gyroid the helicoidal unit cells possess major and minor grooves similar to the DNA double helix structure, and that the major (minor) groove directly connects with the minor (major) groove of the neighboring cell. The D surface, on the contrary, has identical grooves, essentially halving the size of its unit cell in the vertical direction. 

On the other hand, the P surface have horizontally parallel boundary lines and achiral unit cells. We can deform the unit cells so that the its boundary shows a zigzag pattern surrounding the cells, referred to as $\mathrm{P}'$ and shown in Fig.~\ref{fig:fig02}(b). In this way, the $\mathrm{P}'$ cells can be joined with $\mathrm{D}$/$\mathrm{\bar{\reflectbox{D}}}$ and $\mathrm{G}$/$\mathrm{\bar{\reflectbox{G}}}$ and form the corresponding hybrid surfaces, as shown in Fig.~\ref{fig:fig02}(b). Note that since $\mathrm{P}'$ is still achiral, it has to be matched with pairs of glide-related $\mathrm{D}$
and $\mathrm{G}$ tetragonal cells. This is in contrast to the D/G hybrid where only one handiness cells of each constituent surfaces are used. This is schematically shown in the bottom row of Fig.~\ref{fig:fig02}. We note that there are combinatorially infinitely many ways of shuffling and mix-matching any number of these cells so long as one carefully matches the boundary curves while honoring the opposite handiness requirement for neighboring cells if applicable. This point is revisited in the later section where we discuss grain boundaries between these surfaces.

\subsection*{Triangular generalization of P/D/G}
A direct generalization of the above introduced square lattice/tetragonal strip representation of P, D, and G TPMS is to extend to other regular tilings, i.e. edge-to-edge tessellations of identical regular polygons. This leaves us two possibilities: the regular triangular tiling and the regular hexagonal tiling (honeycomb lattice). We provide an example using the triangular tiling in this section that is a direct generalization of gyroid.
As in the case of P/D/G mixtures there are infinitely many possibilities concerning the combination of these other two regular tilings and the corresponding triangular and hexagonal strips. We also note that for the hexagonal tiling due to the handiness of helicoidal strips, there is an additional layer of complexity thanks to geometric frustration, a physical phenomenon often seen in magnetic systems. This will be explored in a different setting in the following section.

\begin{figure}
    \centering
    \includegraphics[width=0.7\linewidth]{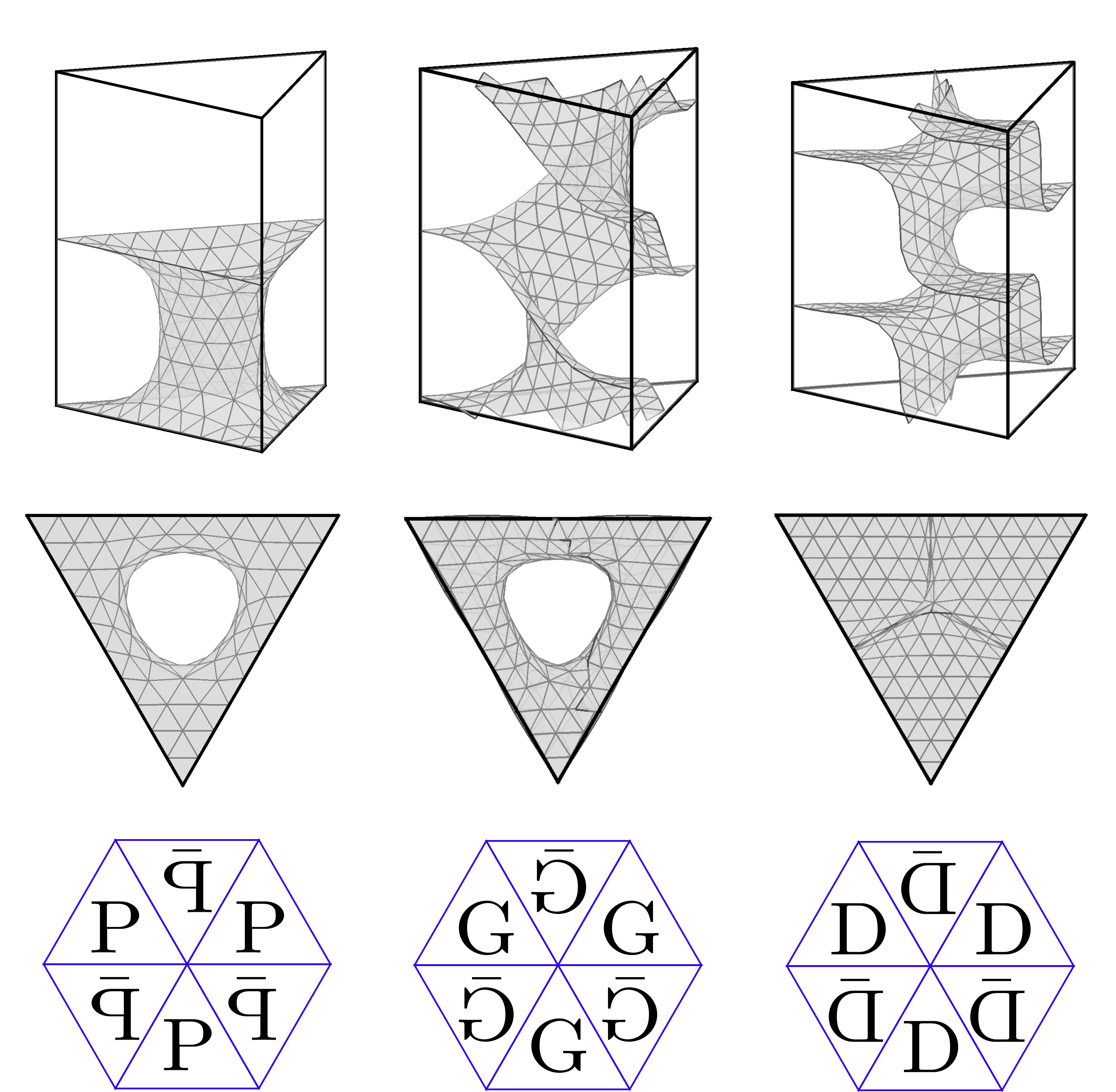}
    \caption{Generalization to triangular strips and tilings.}
    \label{fig:Triangle}
\end{figure}

In Fig.\ref{fig:Triangle} we show the triangular versions of the P/D/G strips and the corresponding surfaces. Generally the composition of triangular P/D/G surfaces follows the same principles as their rectangular counterparts: neighboring tiles are related by glide symmetry planes, flipping the handedness and displacing the periodicity in the $z$-direction by half a period. While the triangular P (the Schwarz's H surface) and D surfaces are different from the original surfaces, it is interesting that the triangular G surface is homeomorphic to the original gyroid. Also, mixtures and grain boundaries of these surfaces can also be derived intuitively again similar to the rectangular cases (not shown).\cite{Chuang2010}

\subsection*{Double diamond}
Generalization of this method is not limited to the change of underlying tilings. In particular, there is an interesting extension for the tetragonal strip of gyroid that is not possible for those of P and D surfaces. Due to the off-center helicoid nature of gyroid strips, one can in principle construct multi-stranded strips by adjusting the ratio between the pitch angle and the girth of the helicoid. For the case of doubly stranded helicoid, shown in Fig.XX, the resulting surface is the double diamond surface provided that we adopt a slightly different tiling rule: the neighboring square tiles are related by plain reflection instead of glide symmetry. This naturally leads to a pair of tetravalent networks each identical to the skeletal graph of D surface.

\begin{figure}
    \centering
    \includegraphics[width=0.9\linewidth]{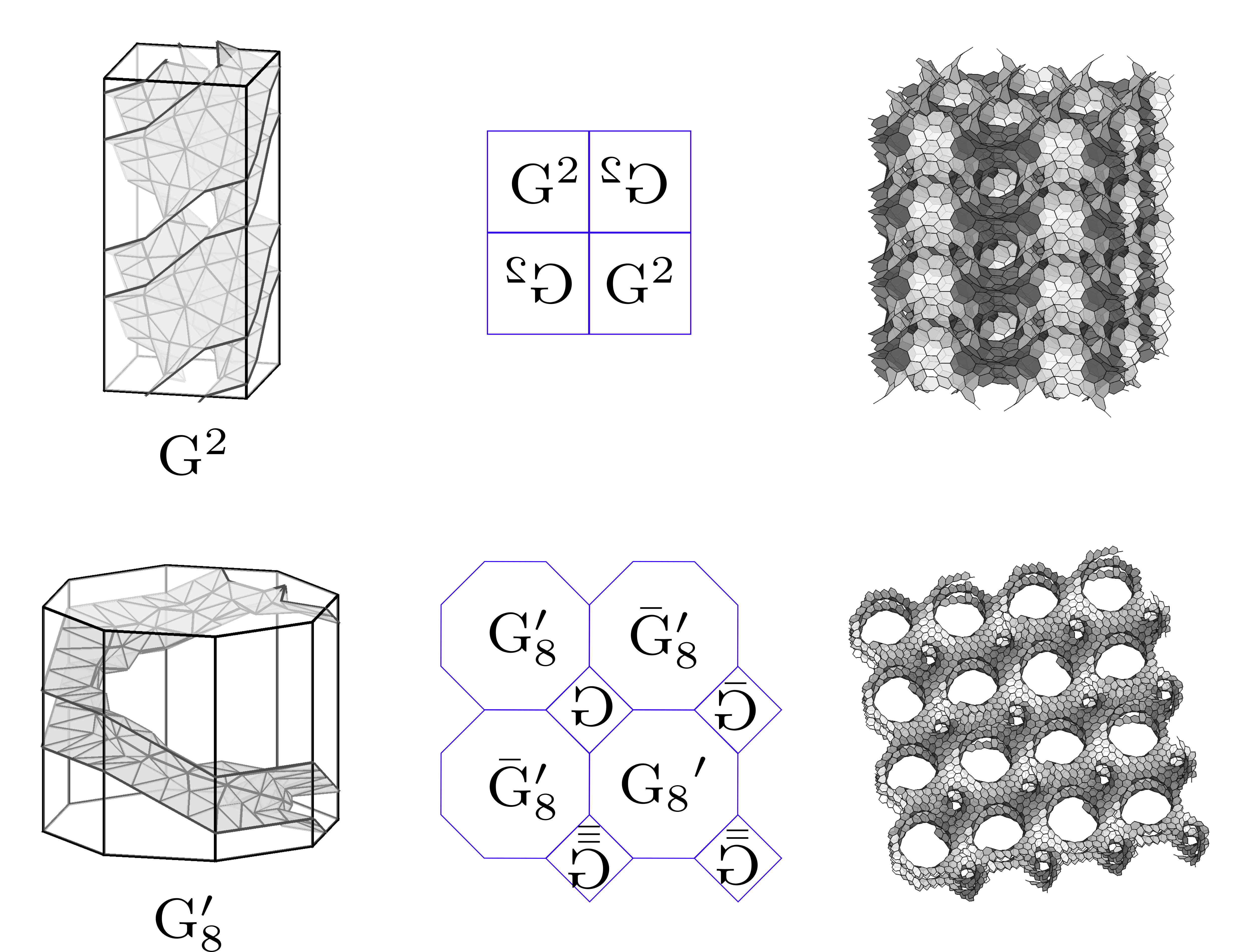}
    \caption{Doubly stranded G strip (upper left) and the double diamond network (upper right) and generalization to the truncated square tiling and the double gyroid network (lower row). A distorted octagonal G strip that is necessary to avoid geometric frustration in this case is shown on the lower left. }
    \label{fig:Double}
\end{figure}

\subsection*{Double gyroid}
To further demonstrate the usefulness of this decomposition-recomposition methodology, we give two additional exemplar hybrid surfaces that are of general interest. The first example is the well known double gyroid structure, two inter-penetrating chiral networks with opposite handiness with a conventional gyroid surface separating them. These networks were described by Wells as (10,3)-a in his pioneering work.\cite{AFWells1977} We show that each one of the chiral networks can be constructed by using a similar procedure described for the P/D/G surfaces generalized for a truncated square lattice, i.e. a Euclidean tiling of $4.8^2$ (one square and two regular octagon per vertex) using the notation of Cundy and Rollett.\cite{CundyRollett} Here, similar to those used in the G surface, one needs a tetragonal helicoidal strip to fill in the squares in the $4.8^2$ tiling. However, similar to the case of double diamond surface, a slight modification to the ratio between the pitch angle and the girth of the helicoid is necessary, making the helicoid ``thinner'' so the resulting network stays in one of the two gyroidal subspace. 

In addition to the tetragonal strip, we also need an octagonal one. Here an extra layer of complexity needs to be accounted for: Suppose that the tetragonal strips are all of the same handedness and since all octagonal tiles share edges with the tetragonal ones, the octagonal strips need to be of the opposite handedness. This immediately leads to conflicting boundary conditions as one is forced to have neighboring octagonal tiles of the same handedness sharing an edge. This geometric frustration is general to all tilings with vertices of odd valencies, provided all polygonal strips are handed. 

To resolve this issue, we modify the octagonal strips so that the pitch angle alternates between zero and a constant finite value that is the negative of that of the tetragonal strips. Essentially, at the edges shared by octagonal tiles the helicoid is bent flat and perpendicular to the $z$-axis. The resulting surface is homeomorphic to one of the two networks of a double gyroid. Naturally the other one can be obtained by a glide symmetry at $x+y=\frac{1+\sqrt{2}}{2}a$, where $a$ is the side length of the tiles. The result is shown in Fig.~\ref{fig:Double}. Notice that due to the tetravalent nature of the corresponding graph, the four tetragonal strips at the corners of an octagonal one are progressively related by one fourth of a full lattice spacing in the $z$ direction, represented by the number of bars. 

\section*{Grain boundaries}
In this section we discuss possible grain boundaries between semi-infinite P/D/G phases by utilizing their tetragonal representations introduced above. We stress that these examples of grain boundaries are certainly not unique. However, as they derive naturally from the tetragonal strip representation, they may add intuition to analyzing these inherently nontrivial construct. 

\subsection*{Horizontal boundaries}
\begin{figure}
    \centering
    \includegraphics[width=0.9\linewidth]{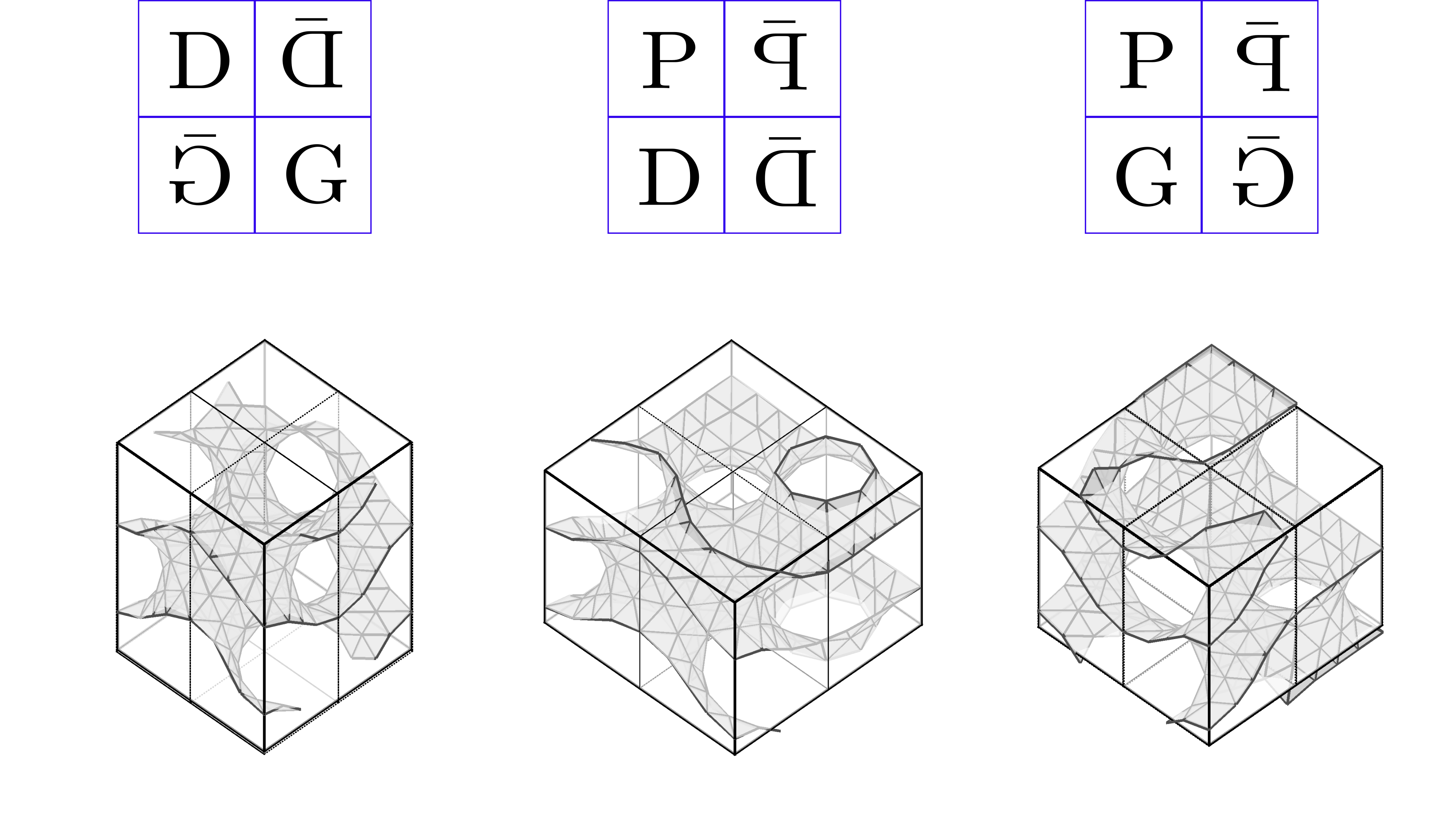}
    \caption{ Grain boundaries between pairs of P/D/G surfaces in the horizontal direction. The 3D renderings are rotated 45 degrees clockwise when viewing from above.}
    \label{fig:Horizontal}
\end{figure}

Following the introduction of P/D/G mixtures in the previous section, it is straightforward to extend our analysis to horizontal grain boundaries between any pair of these TPMS. All three pairs of such grain boundaries are shown in Fig.\ref{fig:Horizontal}. 

\subsection*{Vertical boundaries}
Finally, we turn the attention to grain boundaries in the $z$-direction. As can be seen in the upper row of Fig.~\ref{fig:fig02}, both P and D TPMS have $z$-cross sections in the form of square lattices, albeit in the latter the square lattice is rotated 45 degrees with side length scaled $\sqrt{2}$ times larger than that of P. In light of this, it is straightforward to seamlessly morph a semi-infinite D phase into a P phase that is $\sqrt{2}$ times larger in scale with 45 degrees rotation. 

\begin{figure}
    \centering
    \includegraphics[width=0.9\linewidth]{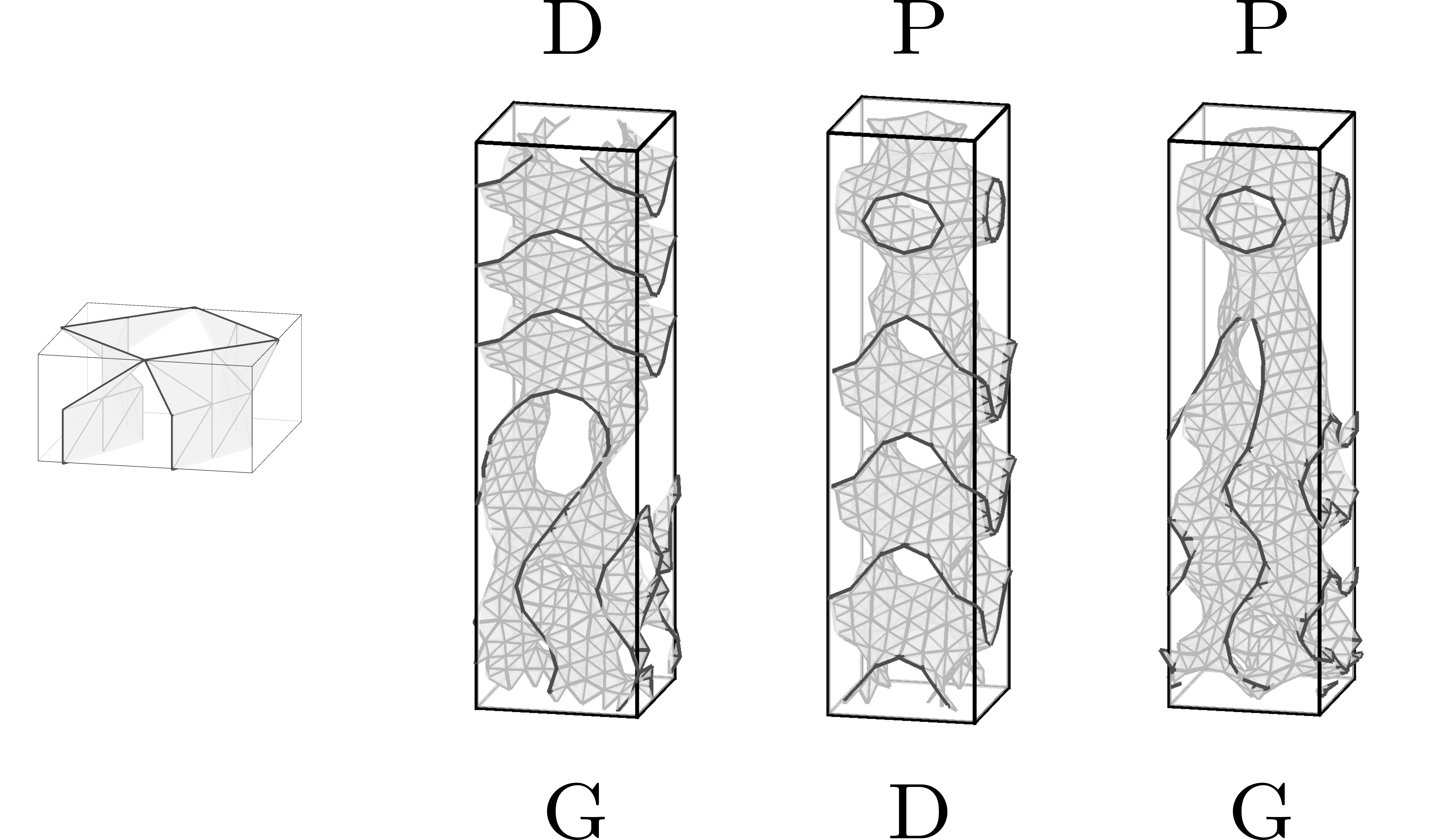}
    \caption{ Grain boundaries between pairs of P/D/G surfaces in the vertical direction. For P/G and D/G cases a transitional layer is needed (left).}
    \label{fig:Vertical}
\end{figure}

The cases involving gyroid are slightly more involved. As shown in Fig.~\ref{fig:fig02}, the $z$-direction cross section is composed of zigzag broken lines. To seamless connect to a P or D counterpart, a transitional layer that distorts the zigzags into a square lattice is necessary. One such layer is shown in Fig.\ref{fig:Vertical}. 

Generally speaking, these boundaries are formed by respecting the \textit{local} geometric matching conditions of the participating surfaces. This implies that there could mixtures of both horizontal and vertical grain boundaries. In principle a 3D island of a P/D/G phase inside of another phase could be constructed, an aspect that might be useful in understanding the nucleation process in the phase transition between these phases in materials. 

\section*{Conclusion}
We introduce a representation for the three canonical triply periodic minimal surfaces (TPMS), the P, D, and gyroid, using a planar square tiling and tetragonal strips that mimic a catenoid (P), a  helicoid (D), and a generalized helicoid (G) surfaces. This allows one to construct and classify a plethora of generalized surfaces that are mixtures of P/D/G or their grain boundaries. By extending to multi-stranded strips as well as other regular (triangular) and semi-regular (truncated square) tilings we provide similar representations to double diamond, double gyroid, and triangular generalizations of P and D surfaces. 

This paper does not provide an exhaustive list of such representation to known TPMS and the area minimality or zero mean curvature conditions are not verified for the ones that are covered in this report, which is left for future work. 
\bibliographystyle{unsrt}
\bibliography{PDGref}

\end{document}